\documentstyle[aps,version2,preprint,epsf]{revtex}
\begin{document}
\draft

\begin{title}
Stochastic to Deterministic Crossover of Fractal Dimensions\\
for a Langevin Equation
\end{title}

\author{
David~A.\ Egolf\\
and\\
Henry~S.\ Greenside\cite{cps_note}
}
\begin{instit}
Department of Physics and\\
Center for Nonlinear and Complex Systems\\
Duke University, Durham, NC 27706\\
dae@phy.duke.edu; hsg@cs.duke.edu
\end{instit}

\begin{abstract}

Using algorithms of Higuchi and of Grassberger and
Procaccia, we study numerically how fractal dimensions
cross over from finite-dimensional Brownian noise at
short time scales to finite values of deterministic
chaos at longer time scales for data generated from a
Langevin equation that has a strange attractor in the
limit of zero noise. Our results suggest that the
crossover occurs at such short time scales that there
is little chance of finite-dimensional Brownian noise
being incorrectly identified as deterministic chaos.

\end{abstract}


\narrowtext

A significant recent effort in nonlinear dynamics has
been devoted to analyzing time series so as to
determine whether they are chaotic or stochastic
\cite{Grassberger91}. Chaotic time series arise from
deterministic strange attractors whose underlying
geometric structure in state space can be used to
improve short term forecasting \cite{Grassberger91}, to
remove noise \cite{Grassberger91}, or to stabilize
unstable periodic orbits by small perturbations of
parameters \cite{Romeiras92}.  Stochastic time series
such as Gaussian white noise or Brownian motion can not
be obtained from any finite set of deterministic
equations, and are less amenable to prediction and
control.

An early expectation was that fractal dimensions of
time series would provide a tool for distinguishing
chaotic from stochastic behavior. A small empirical
fractal dimension would correspond to a deterministic
chaotic time series while a large or diverging fractal
dimension (as a function of embedding dimension) would
correspond to a stochastic time series
\cite{Brandstater83,Berge84}.  This conjecture was
supported by some numerical experiments, e.g., the work
of Ben-Mizrachi et al \cite{BenMizrachi84} which showed
a crossover between high-dimensional white noise and
low-dimensional chaos for time series generated by a
chaotic map with additive white noise. The crossover
occurred as a function of distance between points in
phase space, with large correlation dimensions~$\nu$
(increasing linearly with the embedding dimension)
occurring at small distances and a finite fractal
dimension (independent of embedding dimension)
occurring at larger distances.

Several recent papers
\cite{Osborne89,Theiler91,Greis91} have pointed out
that this conjecture---that a fractal dimension can
distinguish deterministic from stochastic
mechanisms---is not strictly correct.  Osborne and
Provenzale \cite{Osborne89} showed empirically that
stochastic time series with a power-law power spectrum,
$P(\omega) \propto \omega^\alpha$, yield a finite
fractal dimension~$D$ that is parameterized by the
index~$\alpha$; the precise value of~$D$ depends on
some details of the random phases used in generating
the time series \cite{Greis91}.  Theiler has pointed
out that such time series are not likely to be
identified incorrectly as deterministic if observed in
experimental data \cite{Theiler91}. This follows since
time series with a power-law power spectrum have long
correlation times that scale with the length of the
time series, and so are easily identified in the
Grassberger-Procaccia dimension algorithm
\cite{Grassberger83} by rejecting points that are close
in phase space and also close in time \cite{Theiler86}.

In this Brief Report, we explore further the extent to
which stochastic Brownian motions can be distinguished
from chaotic behavior using dimension calculations. Our
starting point is the observation that the calculation
of Mizrachi et al does not include the important case
of a noisy dynamical system that evolves {\sl continuously\/}
in time such as a Langevin equation. Unlike a
noisy map, the paths generated by noisy vector Langevin
equations have a finite fractal dimension in the limit
of short-time scales, with Hausdorff dimension~$D=2$
\cite{Osborne89,LLosa90}.  Over longer but still finite
time scales (long compared to the correlation time of
the time series), low-amplitude noise simply blurs the
structure of the strange attractor and a finite fractal
dimension is found, close to the dimension of the
unperturbed attractor.

There is then the possibility of an interesting
crossover from finite-dimensional stochastic behavior
to finite-dimensional deterministic behavior for
continuous-time noisy systems. In the following, we
study this crossover numerically and find that it
occurs at such short time scales (for fixed moderate
amounts of noise) that it is unlikely that the two
effects will be confused in empirical data. The
results below also confirm some scaling predictions
made by Theiler in his analysis of fractional Brownian
motions \cite{Theiler91}.

In our numerical experiments, we focus our attention on
a vector Langevin equation of the form:
\begin{equation}
  {\bf X}'(t) = {\bf F}( {\bf X} )
   +  \epsilon {\bf G}(t)
  , \label{langevin_eq}
\end{equation}
where the vector field~$\bf F(X)$ is smooth and where
the vector~$\bf G$ represents a delta-function
correlated Gaussian white noise source, $\langle G_i(t)
G_j(t') \rangle = \delta_{i,j} \delta(t - t')$.  In the
limit of large noise strength~$\epsilon \gg 1$, we can
ignore the field~$\bf F$ and we obtain a Brownian-noise
process whose fractal dimension is~$D=2$
\cite{LLosa90}. In the opposing limit of small noise
strength~$\epsilon \ll 1$, the Langevin equation
reduces to a deterministic set of ordinary differential
equations.  To study a possible crossover in fractal
dimension as the noise strength~$\epsilon$ is varied,
we choose the vector field~$\bf F$ to be a
four-dimensional chaotic model derived by Lorenz
\cite{Lorenz84}
\begin{mathletters}
  \begin{eqnarray}
    F_1  &= & X_2 ( X_3 - X_4) - X_1 + c  , \\
    F_2  &= & X_3 ( X_4 - X_1) - X_2 + c  , \\
    F_3  &= & X_4 ( X_1 - X_2) - X_3 + c  , \\
    F_4  &= & X_1 ( X_2 - X_3) - X_4 + c  .
  \end{eqnarray}
\end{mathletters}
For the parameter value~$c=100$, these equations have
chaotic solutions with a fractal dimension~$D \approx
3.3$ \cite{Lorenz84}. This dimension is distinctly
larger than the value for Brownian paths.

We expect the Langevin equation Eq.~\ref{langevin_eq}
to have at least {\sl four\/} different dimension
scaling regimes in the general case, with the size of
each regimes depending on the noise
strength~$\epsilon$, the length of the time
integration~$T$, and details of the strange attractor.
For zero noise strength $\epsilon=0$, we expect a
scaling regime with dimension~$D=1$ at short time
scales and a regime of~$D=3.3$ at long time scales
compared to the correlation time. The former
corresponds to the locally linear geometry of any
smooth path in phase space. When noise is present, a
third scaling regime with~$D=2$ (for Gaussian noise)
should be visible at sufficiently short time scales.
This new regimes arises from the scale-invariant
non-differentiable structure of a Brownian path. For
fixed noise and for sufficiently long time series, a
fourth scaling regime with~$D \to \infty$ should
appear, representing the fact that the path eventually
fills all of space.

We have integrated numerically the Langevin equation
Eq.~\ref{langevin_eq} with the vector field Eq.~2 for
different noise strengths~$\epsilon$ and lengths of
time series~$N$. We used a simple second-order accurate
stochastic Runge-Kutta algorithm \cite{Greenside81},
and have verified that our results are not sensitive to
the choice of time step~$\triangle{t}$. Estimates of
the correlation dimension~$\nu$ were obtained by
delay-embeddings of time series of the first
variable~$X_1(t)$ into spaces of successively higher
dimensions~$e$, with an embedding delay of order the
empirical correlation $\tau_c \approx 0.1$. Although
the state vector~$\bf X$ is known directly from
Eq.~\ref{langevin_eq} so that embedding is not strictly
necessary, scaling regimes are identified more easily
since they are invariant under increasing embedding
dimension~$e$. In our calculations, we also make the
simplifying assumption that there is no observational
noise, a complication that should be examined at a
future date.

Our results are presented in Figures 1--3.  For zero
additive noise~$\epsilon = 0$, Fig.~\ref{pg_dim_n=0}
shows the local slope~$\nu$ (correlation dimension) of
scaling curves obtained with an approximate
Grassberger-Procaccia algorithm, in
which a sample of 1000 points is chosen from the time
series and then {\sl all\/} distances are calculated
between each of these points and every other point
in the entire time series.  The 1000 points were chosen
to be approximately evenly spaced throughout the time series.
This method yielded similar results to Theiler's
box-assisted method \cite{Theiler87},
in which {\sl all\/} distances between points in phase
space are calculated for points that are sufficiently
close. We do not implement Theiler's idea
\cite{Theiler86} of throwing out state-space points
that are close in time, since these points give
valuable information about possible stochastic
structure. The scaling curves are presented over three
different ranges in the phase-space distance~$\tau$
since a single calculation could not span the various
scaling regimes. For quite short time scales $\Delta{t}
= 10^{-9}$, the correlation dimension~$\nu$ is
effectively one, confirming the locally linear smooth
phase-space path.  For larger sampling rates~$\Delta{t}
= 10^{-5}$, one sees a transition from~$\nu = 1$ to
larger values, but there is no apparent scaling
suggestive of the strange attractor. For still larger
sampling rates, now comparable with the typical time
scales of the attractor, one sees a convergence of the
local slope for different embedding dimensions around
the known value~$D \approx 3.3$. This scaling regime is
rather small, spanning just one order of magnitude
in~$\tau$.

Similar calculations at the noise strength $\epsilon =
1.0$ are given in Fig.~\ref{pg_dim_n=1}. There is a new
scaling regime with~$\nu \approx 2.0$ at the smallest
time scales in panel~(a). This indicates the finite
fractal dimension of Brownian motion. This scaling
regime changes smoothly onto the one of
Fig.~\ref{pg_dim_n=0}(a), of locally linear orbits
with~$\nu = 1$. At still larger time scales, we see
that the correlation dimension no longer converges for
different embedding dimensions~$e$, which is the
hallmark of a higher dimensional process.

The most important point of these two figures is that
the crossover from Brownian noise to deterministic
behavior occurs at quite small time scales, and at
quite small distances~$\tau$ between points in phase
space. This conclusion is qualitative in that the size
of the scaling regimes will generally not be known {\sl
a priori\/}; the sizes will depend nontrivially on
the noise strength~$\epsilon$ and on the length of the
time series~$N$.

The difficulty of identifying all the relevant scaling
regimes in dimension plots from time series sampled
with a fixed time step motivated us to explore
alternative ways to see this same structure. Towards
this goal, we used an algorithm given by Higuchi
\cite{Higuchi88_90,Greis91} that estimates a
length-based fractal dimension~$D$ of the graph~$x(t)$
of a curve. This dimension is more appropriate for
stochastic, rather than deterministic, series with
continuous curves giving a value~$D=1$, and
self-similar stochastic curves giving a value~$D=1.5$,
which is the value expected for Brownian motion
\cite{Feder88}.  For a chaotic time series and sampling
rates much longer than the correlation time we expect
and find that $D=2$ since the time series acts as a
plane-filling curve. (We will discuss this point further
in a future publication.)

For three different noise strengths~$\epsilon$,
Fig.~\ref{higuchi_dimensions} gives the local slope
(Higuchi dimension~$D$) of scaling relations as a
function of time interval~$\tau$; the symbol~$\tau$ is
not quite the same as that used in the
Grassberger-Procaccia plots and here directly means a
time interval, rather than a space-state interval.   In
Fig.~\ref{higuchi_dimensions}(a), we see a result that
is equivalent to that already observed in
Fig.~\ref{pg_dim_n=0}(a), namely that there is a smooth
functional dependence at small time scales, for
which~$D=1$, and this crosses over quite clearly to a
chaotic regime of~$D=2$ at time intervals of order the
characteristic time of the attractor.

At moderate noise strengths such as $\epsilon = 0.01$,
there is a new scaling regime at the smallest time
scales with~$D=1.5$, arising from the Brownian motion
of the Langevin path. This crosses over to~$D=1$ at
longer time scales and then to~$D=2$ at still longer
time scales, indicating the different kinds of dynamic
structure present. For still larger noise strengths
such as~$\epsilon = 1.0$, the scaling regime for
Brownian noise increases at the expense of the linear
regime.  For extremely large noise strengths $\epsilon
\gg 1$, the deterministic structure becomes a small
perturbation of the Brownian noise and only the~$D=1.5$
scaling regime should be observable.

{}From the results summarized in these figures, we
suggest that continuous-time processes arising from
nonlinear Langevin equations---a case that models
dynamical experiments perturbed by external
noise---lead to easily separated fractal-dimension
scaling regimes corresponding to stochastic Brownian
motion at short time scales and deterministic chaos at
longer time scales.  This generalizes the earlier
calculations of Mizrachi et al and of Provenzale and
Osborne.

\acknowledgements
This research was supported by NSF grant ASC-8820327,
and by an allotment of CRAY~YMP time through the North
Carolina Supercomputing Center. We are grateful to
James Theiler for providing a copy of his
box-accelerated dimension code, and to NCSA for making
public their HDF libraries and visualization tools.

\figure{
Plots of the Grassberger-Procaccia dimension~$\nu$ for
a fixed zero noise strength $\epsilon=0$, and for three
sampling rates of time series: (a) $\triangle{t} =
10^{-9}$; (b) $\triangle{t} = 10^{-5}$; (c)
$\triangle{t} = 10^{-2}$.  The parameter~$\tau$ here
indicates a distance in a~$e$-dimensional embedding
space. For each sampling rate, a series of length $N =
10^5$ was studied. The curves denoted by squares,
crosses, and triangles correspond to embedding
dimensions of $e = 4$, $e=7$, and~$e=10$ respectively.
\label{pg_dim_n=0}
}

\figure{
The same as for Fig.~\ref{pg_dim_n=0} except the noise
strength has a finite value~$\epsilon=1$. There is a
scaling regime in (a) around~$D=2$ which is typical of
Brownian noise, while the deterministic scaling regime
in (c) at larger time scales is no longer well-defined.
\label{pg_dim_n=1}
}

\figure{
Plots of the Higuchi dimension~$D$ for numerical time
series of the Langevin equation, Eq.~\ref{langevin_eq},
for noise strengths~$\epsilon$ of: (a) $\epsilon =
0.0$; (b) $\epsilon = 0.01$; and (c), $\epsilon = 1.0$.
The parameter~$\tau$ indicates the time interval
separating pairs of points in a time series
\cite{Higuchi88_90}. The curve in each panel is a union of
data obtained from three time series of $10^6$~points
with sampling rates of~$\triangle{t} = 10^{-3},
10^{-6}$, and $10^{-9}$, since a single time series was
not adequate to span all the scaling regimes.
\label{higuchi_dimensions}
}

\pagestyle{empty}
\epsfbox{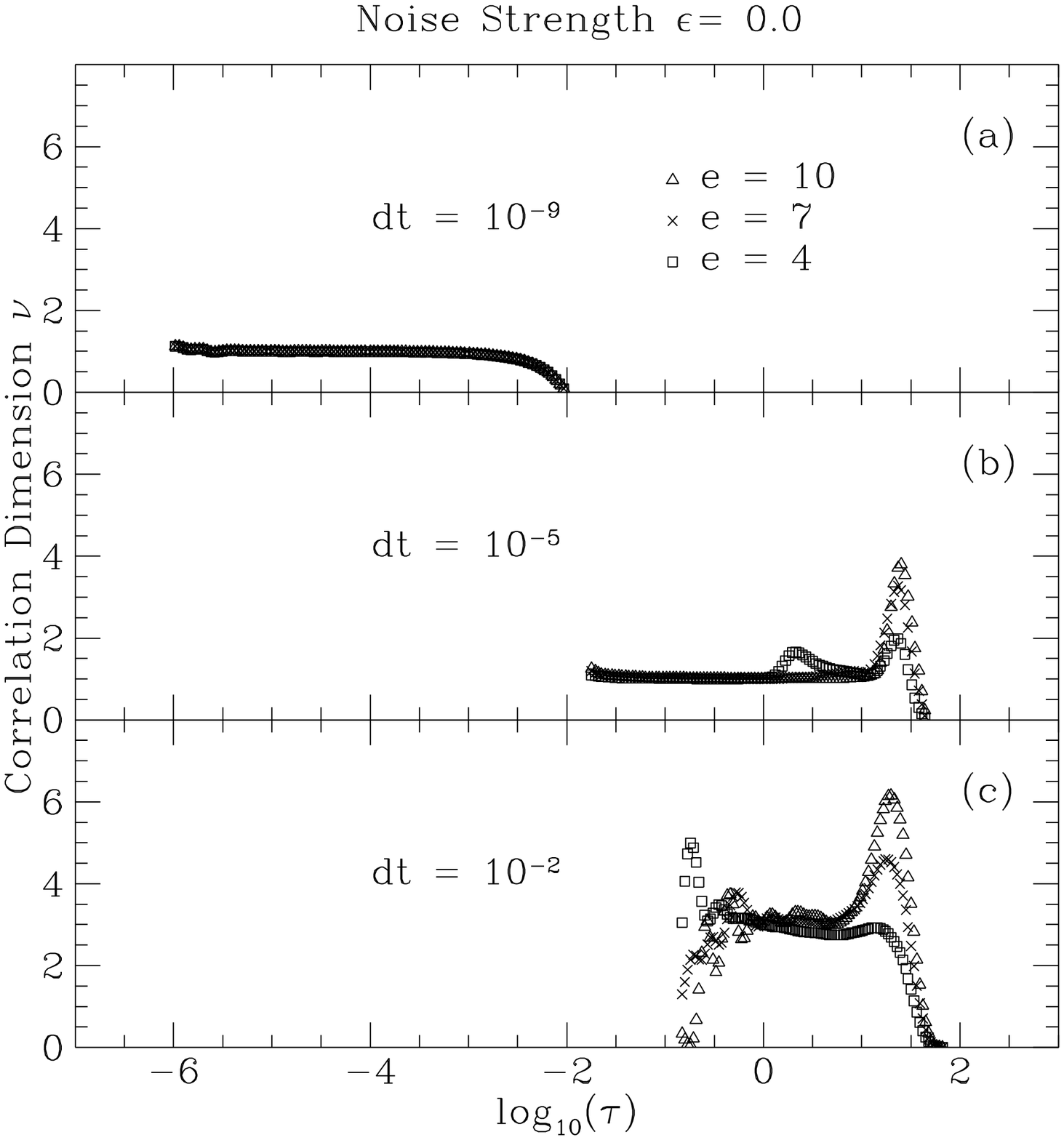}
\vspace*{-0.9in}
\noindent
\baselineskip0pt
{\bf Figure~\ref{pg_dim_n=0}} David A. Egolf
and Henry S. Greenside, Phys.Rev.E,
``Stochastic to Deterministic Crossover of Fractal
Dimensions for a Langevin Equation''

\epsfbox{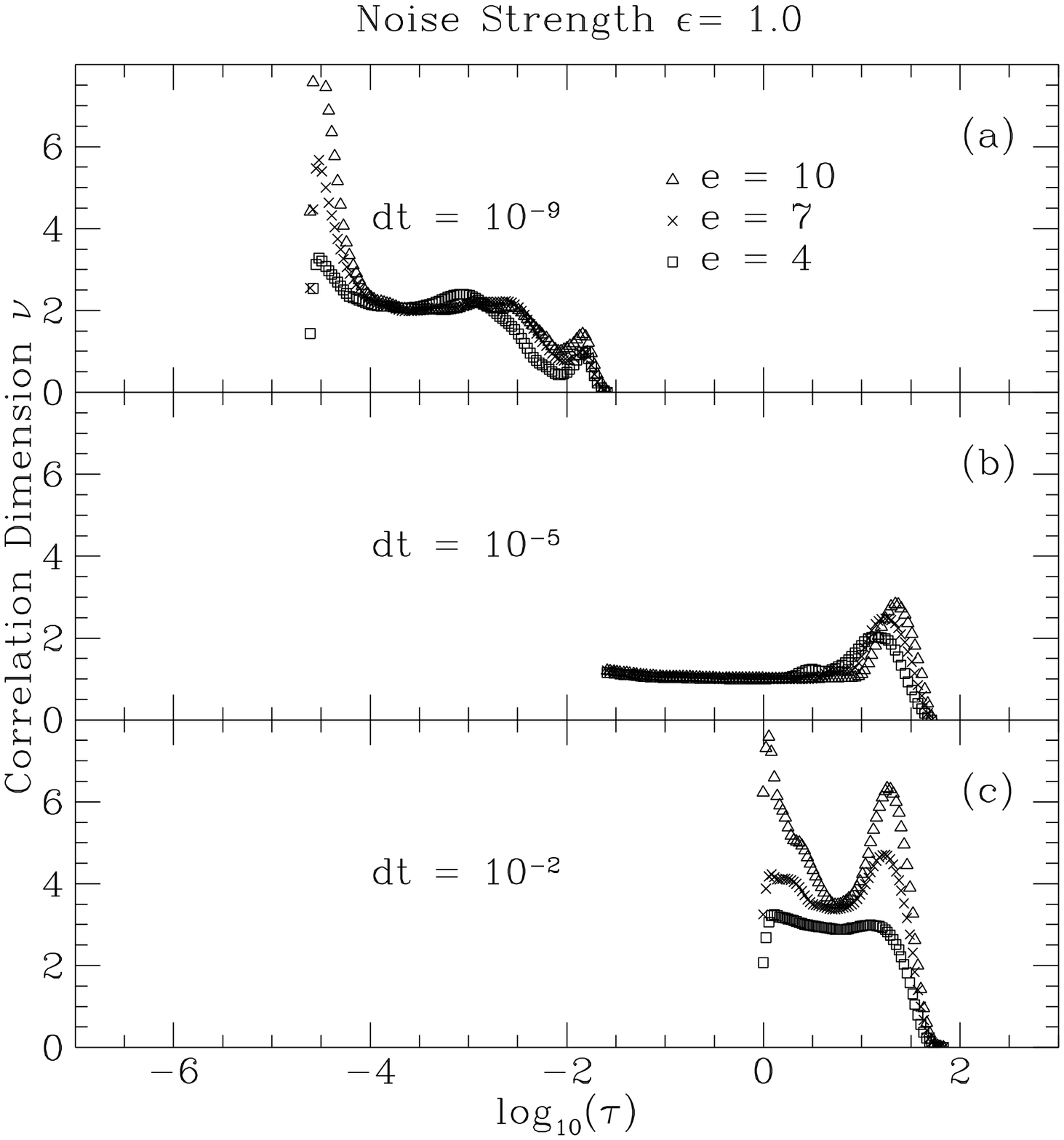}
\vspace*{-0.9in}
\noindent
\baselineskip0pt
{\bf Figure~\ref{pg_dim_n=1}} David A. Egolf
and Henry S. Greenside, Phys.Rev.E,
``Stochastic to Deterministic Crossover of Fractal
Dimensions for a Langevin Equation''

\epsfbox{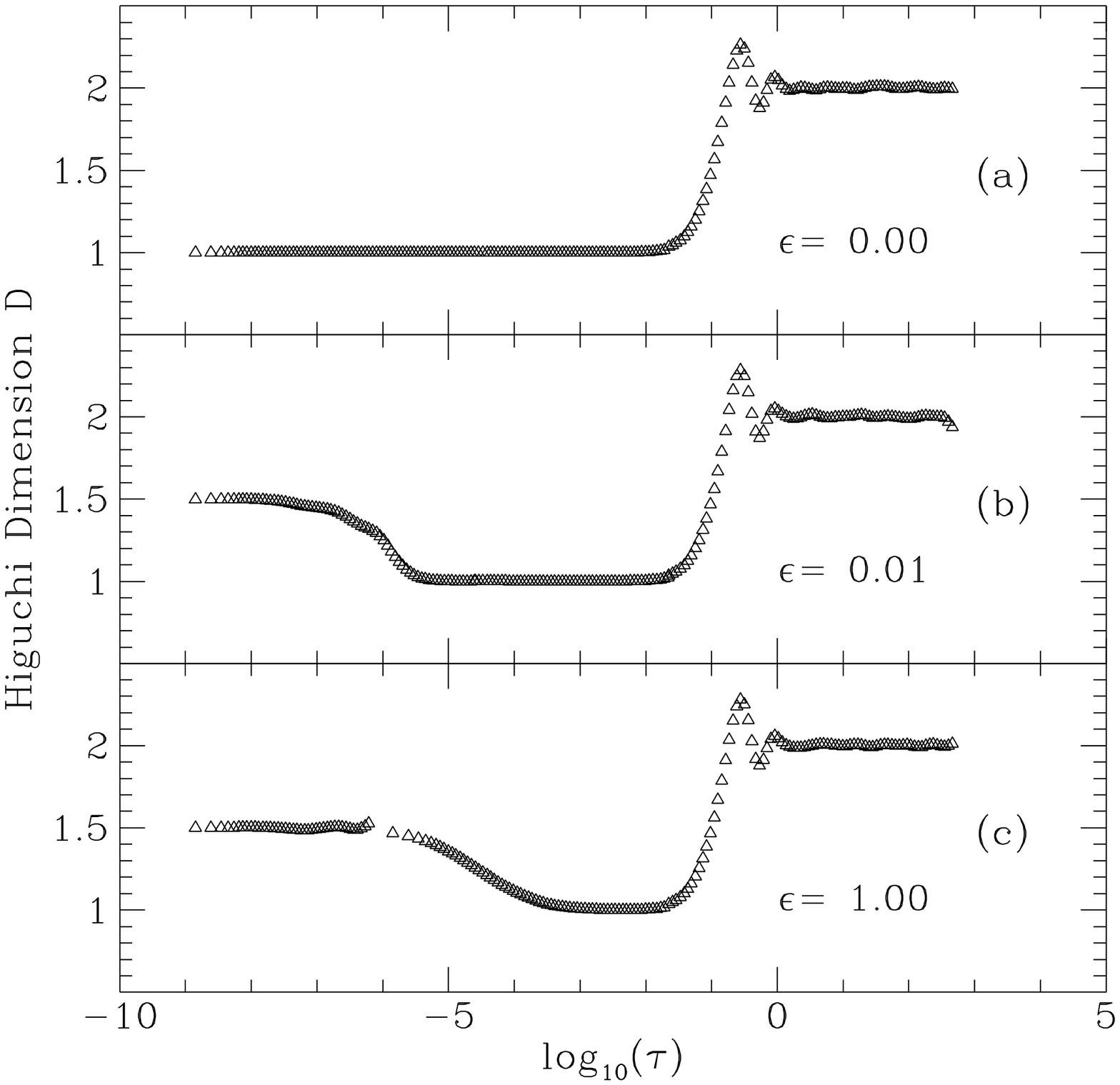}
\vspace*{-0.9in}
\noindent
\baselineskip0pt
{\bf Figure~\ref{higuchi_dimensions}} David A. Egolf
and Henry S. Greenside, Phys.Rev.E,
``Stochastic to Deterministic Crossover of Fractal
Dimensions for a Langevin Equation''

\end{document}